\documentstyle[prc,aps,epsf,psfig,amssymb]{revtex}

\newcommand \be{\begin{eqnarray}}
\newcommand \ee{\end{eqnarray}}
\tightenlines
\begin{document}
\title{Beyond the Quasi-Particle picture in Nuclear Matter calculations
using Green's function techniques}
\author{H. S. K\"ohler\\}
\address{Physics Department, University of Arizona, Tucson, Arizona
85721}
\date{\today}
\maketitle
\begin{abstract}
Widths of low-lying states in nuclei are of the order of 30 MeV. 
These large widths are a consequence of 
the strong interactions leading to a strongly
correlated many body system at the typical densities of nuclear matter.
Nevertheless "traditional" 
Brueckner calculations treat these states as
quasiparticles i.e. with spectral functions of zero widths. 
The width is related to the imaginary part of the selfenergy and is
included selfconsistently in an extension of the Brueckner theory using 
$T$-matrix and Green's function (GF) techniques. 
A more general formulation applicable also to non-equilibrium systems
is contained in the Kadanoff-Baym (KB) equations while still maintaining the
basic many-body techniques of Brueckner theory. In the present work 
the two-time KB-equations are timestepped
along the imaginary time-axis to calculate the binding energy of nuclear
matter as a function of density, including the spectral widths
self-consistently. These zero temperature calculations are compared
with quasi-particle calculations. The inclusion of the selfconsistent
widths are found to add several MeV to the binding.
The spectral widths are due to the long-ranged correlations. Short
ranged correlations decrease rather than increase the binding.
The method is easily 
extended to non-zero temperatures where the importance of the widths are
expected to increase.

\end{abstract}

\section{Introduction\protect\\} 
Nuclei are strongly correlated many-body systems. 
Traditional many-body
theories such as Brueckner's are based on a quasi-particle picture ab
initio
neglecting the widths associated with the strong interactions when calculating
the binding energy. The spectral functions are approximated by
$\delta-$functions.  The Green's function approach with in-medium
T-matrix interactions extend the traditional approach by including the
spectral-functions
selfconsistently.   The calculations work with complex
rather than only real self-energies.\cite{muth03,boz99,boz02} In transport
theories this is somewhat analogous to replacing the Boltzmann equation with
the Kadanoff-Baym (KB) equation. \cite{kad62}

\begin{figure}
\centerline{
\psfig{figure=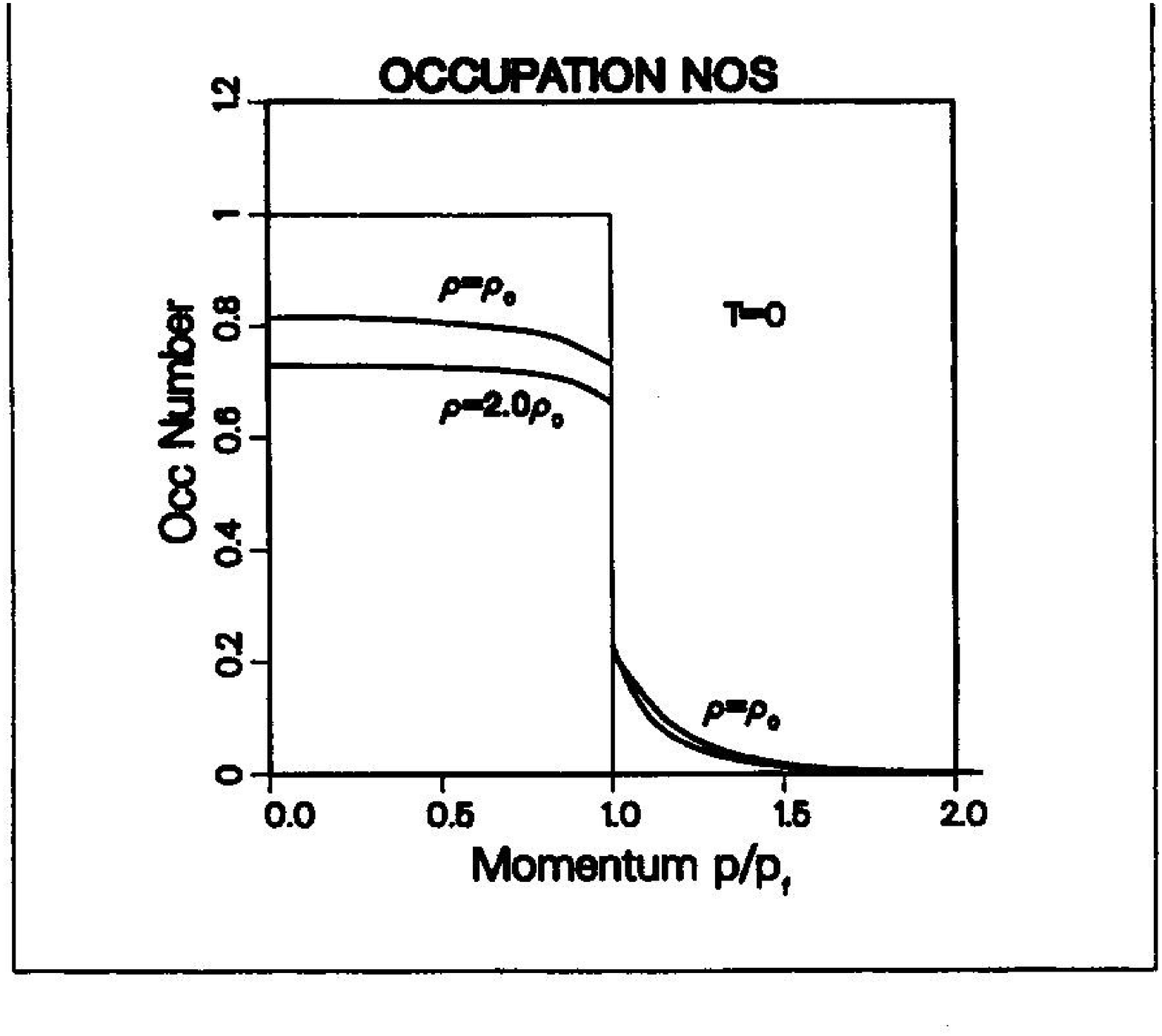,width=8cm,angle=0}
}
\caption{
Occupation numbers in the correlated medium at normal and twice the
normal nuclear matter density as indicated by the thick curves. The
thinner curve shows the occupation numbers for the uncorrelated
(quasi-particle) system.
}
\label{kielfig1}
\end{figure}

The nuclear matter correlations are well
illustrated by regarding  the occupation-numbers in Fig \ref{kielfig1}.
\cite{hsk92}.

In the quasiparticle approximation the occupation-numbers are that of the
fermi-distribution at zero temperature. The correlated medium shows a
depletion of about $20 \%$ at normal nuclear matter density and a number
conserving tail and there exists some experimental evidence of this
"redistribution".

The depletion is mainly due to the short ranged part of the interaction
while as shown in ref \cite{hsk66}, the long-ranged part, defined by the
Moszkowski-Scott (MS) separation method \cite{mos60}, is mainly 
responsible for the widths i.e. the imaginary part of the selfenergy.
This statement is in agreement with the extended quasiparticle 
approximation (EQP)
for the spectral-function given by \cite{kra86,hsk92a,hsk93} 
 \begin{equation}
\ S_{EQP}({\bf p},\omega)=2\pi\hbar\delta(\omega-\omega_{0})
  Z'({\bf p})
-{\cal P}{2\hbar Im\Sigma^{+}({\bf p},\omega) 
\over{(\omega-\omega_{0})^{2}}}
\label{eqp}
\end{equation} 
where
\begin{equation}
 Z'({\bf p})=
  1+({\partial Re\Sigma^{+}({\bf p},\omega)\over{\partial \omega}})
_{\omega=\omega_{0}}
\label{eqZ}
\end{equation}

In the first term the factor $Z'$ that shows the depletion of the
strength depends only on  (the $\omega$-dependence of ) the real part of 
the selfenergy while the width i.e. the imaginary part of the selfenergy
is contained in the second term.

It is of interest to find the effect of 
these widths in a many-body calculation of nuclear properties,
e.g. the binding energy  and saturation of nuclear matter. This can 
conveniently be done using Green's function techniques allowing for complex
self-energies including the widths selfconsistently, going beyond the
quasi-particle picture.\cite{muth03,boz99,boz02}

The effect of correlations on the problems of transport going beyond the
Boltzmann equation by using the KB-equation has been investigated
\cite{dan84,hsk95} and is
of current interest.\cite{here} Here the equilibrium solution of these
same
KB-equations is used to find the effect of the widths in nuclear matter
binding energy.

This method based on non-equilibrium GF methods differs then
from the equilibrium methods mainly in that rather than working with
energy- $(\omega-)$ dependent GF's these depend on two times and in the
equilibrium on the difference between the two times with $\omega$ being
the conjugate of this time-difference. 
 
Some earlier results using a not very realistic local gaussian interaction 
and a second order selfenergy showed the effect
of the widths to increase the binding of nuclear matter by 0.5 MeV at
$k_{F}=1.4$ and by 1.5 MeV at $k_{F}=1.8 fm^{-1}$. \cite{hsk04}
A more realistic interaction will be used here.

Group-renormalisation techniques have been used to derive a $V_{low- k}$
interaction that is independent of the various NN-potential 
models used as input
\cite{bog03}.  It is somewhat analogous to the long-ranged part of
the MS separation method \cite{holt04} that considers the radial
dependence of the interaction while the $V_{low-k}$ is displayed in
momentum space.
A cutoff in momentum-space is chosen to be $\Lambda=2
fm^{-1}$.

\begin{figure}
\centerline{
\psfig{figure=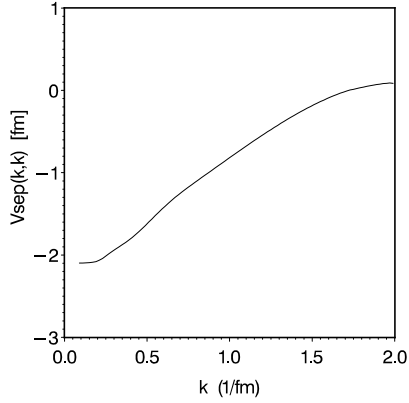,width=6cm,angle=0}
}
\caption{
Diagonal elements of the separable $^{1}S_{0}$ potential for
$\Lambda=2$.
}
\label{vsep1s}
\end{figure}

\begin{figure}
\centerline{
\psfig{figure=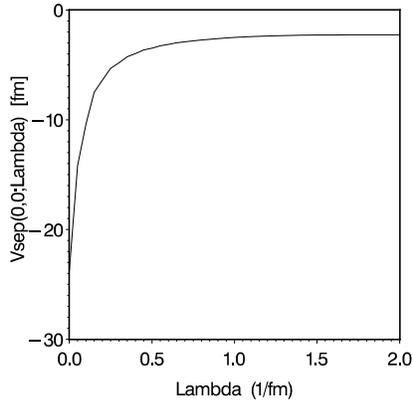,width=6cm,angle=0}
}
\caption{
This figure shows the dependence of $V(\Lambda;0,0)$ on the cutoff  $\Lambda$ 
for the separable $^{1}S_{1}$ potential.
}
\label{vsep}
\end{figure}

A potential model was derived by inverse scattering techniques  \cite{kwo95}
resulting in a separable
NN-potential. The input was experimental scattering phaseshifts and deuteron
data. When used to calculate nuclear matter bindings this potential
showed excellent agreement with the Bonn-potential except in the
$^{3}P_{1}$ case.
In the spirit of the MS-method and the $V_{low-k}$,  
a separable potential $V(\Lambda;k,k')$ 
has been  calculated by inverse scattering
fitting all phaseshifts (exactly) for momenta less than $\Lambda=2 fm^{-1}$.
The diagonal elements of the $^{1}S_{0}$ -potential shown in Fig \ref{vsep1s}
agrees with 
those shown by the authors of $V_{low - k}$.
The off-diagonal elements show a deviation similar
to the spread of the different potential-models considered in the
$V_{low - k}$
publications. The dependence of $V(\Lambda;0,0)$ on the cutoff  $\Lambda$ also
shows perfect agreement especially for the $^{1}S_{0}$ state.(Fig
\ref{vsep})

This serves to show that this separable potential 
also is a good representaion of the $V_{low - k}$ and can be considered
a "realistic" interaction..

\section{Kadanoff-Baym calculations\protect\\} 
The so defined  separable interaction $V(\Lambda=2;k,k')$ 
is used in a Kadanoff-Baym \cite{kad62} formalism.
The earlier
computer-program  used in ref.\cite{hsk04} allowed only for a local
interaction.\cite{hsk99} A revised program allows for 
the separable interaction to
be used. Only S-states are used as these are the major contrbutions to
the widths.
The KB-equations are timestepped along the imaginary
time-axis with selfconsistent (complex) self-energies. With the imaginary time
$\tau$ sufficiently large the resulting correlated GF's correspond to
a temperature $T=0$.\cite{dan84,hsk95} When subsequently time-stepping 
along the real axis the system is stationary and the total energy is calculated
from
\begin{equation}
E= -{i\over{4}}\int{d^{3}{\bf p}\over{(2\pi)^{3}}}({p^{2}\over{2m}}+
i({\partial\over{\partial t}}-{\partial\over{\partial t'}}))
G^{<}({\bf p},t,t')_{t=t'}
\label{energy} 
\end{equation}
which is equivalent to
\begin{equation}
E=4\frac{1}{2}\int_{-\infty}^{+\infty}f(\omega) d\omega \int_{-\infty}
^{+\infty}\frac{d^{3}p}{(2\pi\hbar)^{3}}
[p^{2}/2m-\omega]S({\bf p},\omega)
\label{eq3.1}
\end{equation}
where $S$ is the spectral function and $f$ the fermi-distribution..

In accordance with previous results \cite{hsk66} discussed above the selfenergy can be  calculated from the
long-ranged part of the NN-interaction in 
a second order Born approximation.


The binding energy is
compared with a quasi-particle calculation performed by replacing the
correlated GF's in the KB equations with the free ones which have
zero widths.
\footnote{An extra factor of 1/2 is then required in the expression for
the total energy \cite{fet71,hsk01}}

\begin{figure}
\centerline{
\psfig{figure=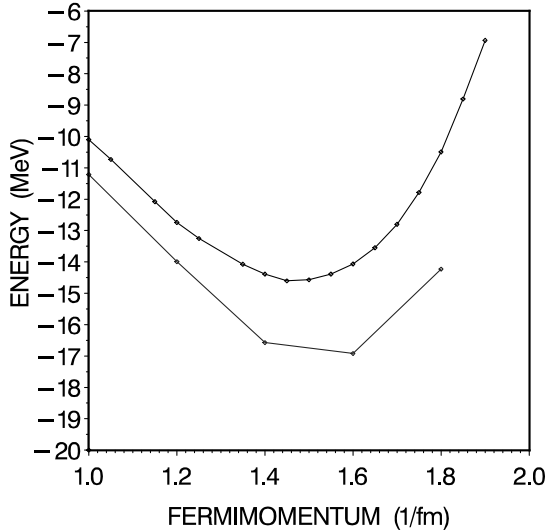,width=8cm,angle=0}
}
\caption{
Upper curves shows the quasiparticle (standard Brueckner) result while
in the lower curve the correction due to correlated GF's i.e widths is
shown.
}
\label{sep2im}
\end{figure}

In
Fig \ref{sep2im} the upper curve shows the result of the quasiparticle
calculation in a "standard" Brueckner calculation with the separable
interaction and the Bonn-B deuteron data as in ref.\cite{kwo95}.
The difference between the
quasi particle and selfonsistent spectral function calculations is added
to get the lower curve thus showing the effect of going beyond the
quasi-particle approximation. 

\section{Conclusions\protect\\} 

Fig \ref{sep2im} shows an increase 
in binding of about 2 MeV due to the widths of the spectral function
relative to the quasi-particle result
while the results of full $T$-matrix calculations
\cite{muth03,boz02} show a decreased rather than an increased binding.
These two results are however not directly comparable.

Various
methods and approximations were discussed at length in refs \cite{hsk92a,hsk93}
where one notable result obtained was  that the binding
calculated with spectral functions gave a binding of $17.4$ MeV while
the Brueckner (quasi-particle) calculation gave $20.0$ MeV. This seems
comparable to the findings in refs\cite{muth03,boz02}.

But the result shown in Fig \ref{sep2im} only includes the effect 
of the width of the spectral function while
in relation to  Fig \ref{kielfig1} and eq ({\ref{eqp}) another 
effect of
the strong correlations in nuclei was pointed out:
The
depletion of occupation-numbers due to short-ranged correlations. 
An estimate of the corrections to the binding energy related to this
depletion can be made as follows.
The depletion is directly related to the
third order "rearrangement" energy that is positive and when included
in the definition of single-particle enegies in the calculation of the
Brueckner $K$-matrix will lead to an increase of about one MeV in the
binding energy. \cite{hsk69,hsk75} An additional correction is
for the reduced
strength of the spectral function in eq (\ref{eq3.1}).
This is given by the factor $Z'$ in eq (\ref{eqp})
shown by the reduced oocupation-numbers  in Fig \ref{kielfig1}
to be $\approx 0.8$ so that the binding due to this correction 
is reduced by $\approx 20\%$ i.e.  $\approx 3$ MeV. 
The total effect associated with the depletion would thus be a 
binding decreased by  $\approx 2 MeV$.
This would thus essentially cancel the effect of the width shown in Fig
\ref{sep2im}.

The above are  estimates that could be off by MeV's
and are only meant to illustrate that there are opposing effects that
each are not small justifying full $T$-matrix
calculations such as referred to above.

These spectral corrections are expected to increase with density and temperature
and would be important for a theoretical determination of 
the Nuclear Equation Of State.
Of course these calculations are still limited by our
limited knowledge of nuclear forces and many body effects.

The method of imaginary time-stepping used here has the numerical
advantage over the conventional $T$-matrix calculations 
of not explicitly having to integrate over sharply peaked
spectral functions as a function of $\omega$.  
For the study of non-equlibrium phenomena the method can also 
be used
with constraints to define an initial condition.\cite{dan84}
It still has to be improved however to incorporate
a full
$T$-matrix calculation rather than just a second order Born calculation.
Only $S$-states were included to find the energy correction due to  the
width. This may not be sufficient.

In summary it can now be conclusively stated that selfconsistent spectral
functions should (and perhaps equally important now \it can \rm) 
be included in any serious consideration of nuclear 
properties being it in ground state, excited state or in a state of
non-equilibrium.

\end{document}